\documentclass[12pt]{article}
\usepackage{graphicx}
\catcode`\@=11
\newcommand{\abs}[1]{\left\vert#1\right\vert}
\newcommand{\bin}[2]{{#1\choose#2}}
\renewcommand{\d}{{\rm d}}
\newcommand{\e}{{\rm e}}
\newcommand{\erf}{\mathop{{\rm erf}}}
\newcommand{\erfc}{\mathop{{\rm erfc}}}
\newcommand{\frad}[2]{\displaystyle{\displaystyle#1\over\displaystyle#2}}
\newcommand{\ii}{{\rm i}}
\newcommand{\integ}{\mathop{{\rm Int}}}
\newcommand{\mean}[1]{\left\langle#1\right\rangle}
\newcommand{\re}{\mathop{{\rm Re}}}

\newcommand{\vb}{{\vphantom{\displaystyle{M}}}}
\renewcommand{\L}{{\cal L}}
\renewcommand{\P}{{\cal P}}
\begin{document}

\centerline{\Large\bf Statistics of the occupation time}
\smallskip
\centerline{\Large\bf for a random walk in the presence of a moving boundary}

\vspace{1cm}

\centerline{\large
C.~Godr\`eche$^{a,}$\footnote{godreche@spec.saclay.cea.fr}
and J.M.~Luck$^{b,}$\footnote{luck@spht.saclay.cea.fr}}
\vspace{1cm}
\centerline{$^a$Service de Physique de l'\'Etat Condens\'e,
CEA Saclay, 91191 Gif-sur-Yvette cedex, France}
\vspace{.1cm}
\centerline{$^b$Service de Physique Th\'eorique\footnote{URA 2306 of CNRS},
CEA Saclay, 91191 Gif-sur-Yvette cedex, France}

\vspace{1cm}

\begin{abstract}
We investigate the distribution of the time spent by a random walker
to the right of a boundary moving with constant velocity $v$.
For the continuous-time problem (Brownian motion),
we provide a simple alternative proof of Newman's recent result
[J.~Phys.~A {\bf 34}, L89 (2001)]
using a method due to Kac.
We then discuss the same problem for the case of a random walk in discrete time
with an arbitrary distribution of steps, taking advantage of the general
set of results due to Sparre Andersen.
For the binomial random walk
we analyse the corrections to the continuum limit
on the example of the mean occupation time.
The case of Cauchy-distributed steps is also studied.
\end{abstract}

\vfill
\noindent To be submitted for publication to Journal of Physics A
\vskip -1pt
\noindent P.A.C.S.: 02.50.Ey, 02.50.Ga, 05.40.+j

\newpage
\setcounter{footnote}{0}
\section{Introduction}

Consider a Brownian particle, starting from the origin, whose position $x_t$
satisfies the Langevin equation
\[
\frac{\d x_{t}}{\d t}=\eta_{t},
\]
where $\eta_{t}$ is Gaussian white noise, such that
$\left\langle\eta_{t}\right\rangle=0$ and
$\left\langle\eta_t\eta_{t'}\right\rangle=2D\,\delta(t-t')$.

In a recent paper~\cite{newman}, Newman considered the following question:
What is the distribution of the
length of time spent by the particle to the right of a boundary moving with
constant velocity $v$?
This quantity, denoted by $T_{t}^{+}$, is known as
the occupation time of the half-line located to the right of the boundary.
It reads
\[
T_{t}^{+}=\int_{0}^{t}\d t'\,I_{t'},\qquad I_{t'}=\Theta(x_{t'}-vt'),
\]
where $\Theta(x)$ is the Heaviside function.
The indicator random variable $I_t$ is therefore equal to $1$ if $x_{t}>vt$,
and $0$ else.
Similarly, the occupation time to the left of the moving boundary
is denoted by $T_{t}^{-}$, and such that $T_{t}^{+}+T_{t}^{-}=t$.

A number of past studies as well as more recent ones have been devoted to
the statistics of the occupation time
of stochastic processes, either in probability
theory~\cite{levy,kac,lamperti,cox},
or in statistical
physics~\cite{dg,baldass,drg,dubna,newman1,newman2,newman3,maj,smedt,glRenew}
in the context of persistence.

A derivation of the probability density of $T_{t}^{+}$,
$f_{T_{t}^{+}}(t,\tau)=\d\P(T_{t}^{+}<\tau)/\d\tau $,
is given in Ref.~\cite{newman}, with the result
\begin{equation}
f_{T_{t}^{+}}(t,\tau)=F^{+}(\tau,v)F^{-}(t-\tau,v),
\label{fT>}
\end{equation}
where
\begin{equation}
F^{\pm}(\tau,v)=
\frac{1}{\sqrt{\pi\tau}}\,\exp\left(-\frad{v^2\tau}{4D}\right)
\mp\frac{v}{2\sqrt{D}}\erfc\left(\pm\frac{v}{2}\sqrt\frac{\tau}{D}
\right),
\label{Fpm}
\end{equation}
and $\erfc$ is the complementary error function.

In particular, in the case of a static boundary~($v=0)$,
we have
\begin{equation}
f_{T_{t}^{+}}(t,\tau)=\frac{1}{\pi}\frac{1}{\sqrt{\tau(t-\tau)}},
\label{static}
\end{equation}
hence the fraction of time $T_{t}^{+}/t$ spent by the Brownian particle to
the right of the origin admits a limiting distribution as $t\rightarrow\infty$,
which reads
\begin{equation}
\lim_{t\rightarrow\infty}f_{T_t^+/t}(x)=\frac{1}{\pi\sqrt{x(1-x)}}
\qquad(0<x<1).
\label{arcsine}
\end{equation}
The arcsine law~\cite{levy,feller} is thus recovered.

The aim of this note is to complement Newman's work in two directions.

Firstly, we give an alternative, simpler derivation
of equations~(\ref{fT>}),~(\ref{Fpm}),
using a method originally due to Kac~\cite{kac}.

Secondly, we discuss the corresponding problem
for a sum of random variables, i.e., for a random walk in discrete time,
with an arbitrary distribution of steps, either narrow or broad.
The general set of results due to Sparre Andersen~\cite{sparre,feller} is the
starting point of
this analysis.
A factorisation property of the distribution of the occupation time
similar to~(\ref{fT>}) holds,
the role of $F^{+}(\tau,v)$ being played by the quantity $F^+_k(v)$,
which is simply the survival probability of the walk in the presence
of the boundary.
We investigate two examples in more detail.
For the binomial random walk,
for which a detailed study of $F^+_k(v)$ can be found in~\cite{bauer},
we analyse
the corrections to the continuum limit
on the example of the mean occupation time;
for a Cauchy distribution of steps, we determine the probability
distribution of the occupation time.

\section{Brownian motion}

The problem of a particle executing symmetric Brownian
motion in the presence of a boundary moving with constant velocity $v$
is equivalent to that of biased Brownian motion with velocity $-v$
in the presence of a fixed boundary, located at the origin.
Consider the probability for this biased Brownian
walk to be at position $x$ at time~$t$, and to have spent a length of time
equal to $\tau$ to the right of the origin.
The joint probability density of the event~($x_t=x$, $T_t^+=\tau$)
is denoted by $p(t,\tau,x)$.
We then have
\begin{equation}
f_{T_{t}^{+}}(t,\tau)=\int_{-\infty}^{\infty}\d x\,p(t,\tau,x).
\label{jint}
\end{equation}
The method of Kac\footnote{This method has recently been applied
to the investigation of the distribution of the occupation time
of subordinated Brownian motion~\cite{maj} (see also~\cite{smedt}).}
consists in writing a master equation for $p(t,\tau,x)$.
It is an easy matter to realize that the latter reads,
in the present case,
\begin{equation}
\frac{\partial p}{\partial t}+\Theta(x)\frac{\partial p}{\partial\tau}
=D\frac{\partial^2p}{\partial x^2}+v\frac{\partial p}{\partial x},
\label{jmas}
\end{equation}
with initial condition $p(0,\tau,x)=\delta(x)\,\delta(\tau)$.
In Laplace space, setting
\[
\hat{p}(s,u,x)=\mathrel{\mathop{\L}\limits_{t,\tau}}\,p(t,\tau,x),
\]
equation~(\ref{jmas}) yields
\begin{equation}
\left(s+u\Theta(x)\right)\hat{p}
-D\frac{\partial^2\hat{p}}{\partial x^2}-v\frac{\partial\hat{p}}{\partial x}
=\delta(x).
\label{jmaslap}
\end{equation}
This inhomogeneous differential equation
is easily solved.
By requiring $\hat{p}(s,u,x)$ to vanish at infinity
($x\rightarrow\pm\infty$), we have
\[
\hat{p}(s,u,x)=A(s,u)\times\left\{\matrix{
\exp\left[-\left(v+\sqrt{v^{2}+4D(s+u)}\right)\frad{x}{2D}\right]
\hfill&(x\ge0),\hfill\cr
\exp{\left[\left(-v+\sqrt{v^{2}+4Ds}\right)\frad{x}{2D}\right]}^\vb
\hfill&(x\le0).\hfill
}\right.
\]
The amplitude $A(s,u)$ is determined by the right side
of~(\ref{jmaslap}), yielding the condition
\[
\frac{\partial p(t,\tau,0^+)}{\partial x}
-\frac{\partial p(t,\tau,0^-)}{\partial x}=-\frac{1}{D},
\]
hence
\[
A(s,u)=\frac{\sqrt{v^{2}+4D(s+u)}-\sqrt{v^{2}+4Ds}}{2Du}.
\]
Finally equation~(\ref{jint}) yields
\begin{equation}
\hat{f}_{T_t^+}(s,u)
=\mathrel{\mathop{\L}\limits_{t}}\left\langle\e^{-uT_t^+}\right\rangle
=\int_{-\infty}^{\infty}\d x\,\hat{p}(s,u,x)
=\hat{F}^+(s+u,v)\hat{F}^-(s,v),
\label{jlap}
\end{equation}
with
\begin{equation}
\hat{F}^{\pm}(u,v)
=\frac{2\sqrt{D}}{\sqrt{v^2+4Du}\pm v}.
\label{Flaplace}
\end{equation}
These functions are precisely the Laplace transforms with respect to $\tau$
of $F^{\pm}(\tau,v)$ given in equation~(\ref{Fpm}).
An inverse Laplace transformation finally leads to the result~(\ref{fT>}).

Let us now comment on the form of the solution~(\ref{fT>}).
A first comment concerns symmetry.
It is clear that, for a given velocity $v$,
$T_{t}^{+}(v)$ has the same distribution as $T_{t}^{-}(-v)$.
Therefore
\[
f_{T_{t}^{+}(-v)}(t,\tau)=f_{T_{t}^{-}(v)}(t,\tau)=f_{T_{t}^{+}(v)}(t,t-\tau).
\]
This requirement is satisfied by the solution~(\ref{fT>}), since
$F^\pm(\tau,-v)=F^\mp(\tau,v)$, according to~(\ref{Fpm}), or
equivalently $\hat F^\pm(u,-v)=\hat F^\mp(u,v)$,
according to~(\ref{Flaplace}).
A second comment is on normalisation.
Integrating equation~(\ref{fT>}) upon $\tau\in[0,t]$ yields
\[
1=\P(T_{t}^{+}<t)=\int_{0}^{t}\d\tau\,f_{T_{t}^{+}}(t,\tau)
=F^{+}(t,v)*F^{-}(t,v),
\]
the star in the right side denoting a convolution product.
Hence in Laplace space the equality $\hat{F}^{+}(s,v)\hat{F}^{-}(s,v)=1/s$
should hold.
This is indeed the case, as can be seen from~(\ref{Flaplace}).

For $v=0$, the distribution of the occupation time is given by~(\ref{static}).
The first correction to this behaviour for small $v$ is given by
$$
f_{T_{t}^{+}}(t,\tau)=\frac{1}{\pi}\frac{1}{\sqrt{\tau(t-\tau)}}
+\frac{v}{2\sqrt{\pi
D}}\left(\frac{1}{\sqrt{\tau}}-\frac{1}{\sqrt{t-\tau}}\right)
+\cdots
$$
As a consequence,
\[
\mean{T_t^+}=\frac{t}{2}\left(1-\frac{2v}{3}\sqrt\frac{t}{\pi D}+\cdots\right),
\]
showing that the presence of any bias $v\ne0$
is relevant in the long-time regime.
There is actually a non-trivial limiting distribution
for $T_t^+$ as $t\to\infty$ if $v>0$ (and, by symmetry, for $T_t^-$ if $v<0$).
We have indeed
$\lim_{t\to\infty}F^-(t,v)=\lim_{s\rightarrow0}\,s\hat{F}^-(s,v)=v/\sqrt{D}$,
so that, by~(\ref{fT>}),
\begin{equation}
f_{T^+}(\tau)
=\frac{v}{\sqrt{D}}\,F^+(\tau,v),
\label{jlim}
\end{equation}
with the notation $T^+=\lim_{t\to\infty}T_t^+$.
Using the asymptotic expansion
$$
\erfc(x)=\frac{\e^{-x^2}}{x\sqrt{\pi}}\left(1-\frac{1}{2x^2}+\cdots\right),
$$
we see that the distribution~(\ref{jlim})
falls off exponentially for large $\tau$, as
\[
f_{T^+}(\tau)\approx\frac{2}{v}\,\sqrt\frac{D}{\pi\tau^3}
\,\exp\left(-\frad{v^2\tau}{4D}\right),
\]
so that all the moments $\mean{(T^+)^n}$ are finite.
The latter can be computed by noting that
$$
\hat f_{T^+}(u)
=\frac{v}{\sqrt{D}}\,\hat F^+(u,v),
$$
which, expanded around $u=0$, leads to
\begin{equation}
\mean{(T^+)^n}=\frac{(2n)!}{(n+1)!}\left(\frac{D}{v^2}\right)^n\qquad(n\ge1).
\label{moment}
\end{equation}

\section{Random walk in discrete time}

Consider a discrete random walk defined by a sum of independent,
identically distributed random variables:
\[
x_n=\sum_{i=1}^n\eta_i,
\]
with an arbitrary distribution of the steps $\eta_i$
(either discrete or continuous, narrow or broad).
For this random walk, the occupation
time to the right of the boundary moving with velocity $v$ is defined as
\[
T_{n}^{+}=\sum_{m=1}^{n}I_{m},\qquad I_{m}=\Theta(x_m-vm),
\]
hence the indicator random variable $I_m=1$ if $x_m>vm$, or $0$ else.
The occupation time $T_{n}^{-}$ to the left of the boundary is defined
likewise, and such that $T_{n}^{+}+T_{n}^{-}=n$.
As above, we note the
equivalence of the problem thus stated with that of the occupation time of a
random walk with biased steps $\eta_{i}-v$
in the presence of a fixed boundary, located at the origin.

For sums of independent random variables,
a result due to Sparre Andersen~\cite{sparre,feller}
expresses the probability distribution of $T_{n}^{+}$ as the product
\begin{equation}
\mathcal{P}(T_{n}^{+}=k)=\mathcal{P}(T_{k}^{+}=k)\mathcal{P}(T_{n-k}^{-}=n-k).
\label{spar1}
\end{equation}
In this equation, $\mathcal{P}(T_{k}^{+}=k)$, hereafter
denoted by $F_{k}^{+}(v)$, is the
probability that the walk remained to the right of the boundary up to
time $k$, or
\[
F_{k}^{+}(v)=\mathcal{P}(T_{k}^{+}=k)=\left\langle I_{1}I_{2}\ldots
I_{k}\right\rangle.
\]
Similarly,
\[
F_{k}^{-}(v)=\mathcal{P}(T_{k}^{-}=k)=\left\langle\left(1-I_{1}\right)
\left(1-I_{2}\right)\ldots\left(1-I_{k}\right)\right\rangle
\]
is the probability that the random walker
remained to the left of the boundary up to time $k$.
In other words, the quantities $F_k^\pm(v)$
are survival probabilities of the walk in the presence of the boundary, 
up to time $k$~\cite{dg,baldass,drg,dubna,bauer,dornic}.
For instance, 
$$
F_{k}^{+}(v)=\mathcal{P}\left ( x_m>vm\hbox{ for } 1\le m\le k\right ).
$$

The generating function of the $F_n^+(v)$
is related to the generating function of the one-time probabilities
$\left\langle I_{n}\right\rangle=\P(x_{n}>vn)$ by~\cite{sparre,feller}
\begin{equation}
\sum_{n=0}^{\infty}F_{n}^{+}(v)z^{n}
=\exp\left(\sum_{n=1}^\infty\frac{z^n}{n}\left\langle I_n\right\rangle\right).
\label{spar2}
\end{equation}
Equation~(\ref{spar1}) is the discrete counterpart of equation~(\ref{fT>}).
Together with~(\ref{spar2}), it provides the answer to the question posed
(in terms of the one-time quantities $\mean{I_n}$).

We now illustrate the above formalism by two examples.
First, for a narrow distribution of steps $\eta_i$,
we determine the continuum limit of equations~(\ref{spar1}),~(\ref{spar2}),
thus recovering the results~(\ref{fT>}),~(\ref{Fpm})
obtained for Brownian motion.
For the binomial random walk, we investigate
the corrections to the continuum limit
on the example of the mean occupation time.
Then, for a Cauchy distribution of steps,
we determine the distribution of the occupation time
$\mathcal{P}(T_{n}^{+}=k)$ explicitly.

Consider a narrow distribution of steps, with $\left\langle\eta\right\rangle=0$
and $\left\langle\eta^2\right\rangle=\sigma^2$.
The continuum limit is defined as $n\rightarrow\infty$, $v\rightarrow 0$,
with $v\sqrt{n}/\sigma=\xi$ fixed.
The central limit theorem yields
\begin{equation}
\mean{I_n}=\P(x_{n}>vn)
=\P\left(\frac{x_{n}}{\sigma\sqrt{n}}>\xi\right)
\approx\frac{1}{\sqrt{2\pi}}\int_{-\infty}^{\xi}\d u\,\e^{-u^2/2}
=\frac{1}{2}\left(1-\erf\frac{\xi}{\sqrt{2}}\right).
\label{jclt}
\end{equation}

Let us analyse equation~(\ref{spar2}) in the same limit.
Setting $z=\e^{-s}$, to leading order as $s\to 0$,
identifying generating series with Laplace transforms,
we obtain the following estimates:
\begin{eqnarray*}
&&\sum_{n=0}^{\infty}F_{n}^{+}(v)z^{n}\approx\hat{F}^{+}(s,v),\qquad
\sum_{n=1}^{\infty}\frac{z^{n}}{n}=-\ln(1-z)\approx -\ln s,\\
&&\sum_{n=1}^{\infty}\frac{z^{n}}{n}\erf\frac{v\sqrt{n}}{\sigma\sqrt{2}}
\approx\int_{0}^{\infty}\frac{\d t}{t}\e^{-st}\erf\frac{v\sqrt{t}}
{\sigma\sqrt{2}}=2\ln\left(\frac{v}{\sigma\sqrt{2s}}+\sqrt{\frac{v^2}
{2\sigma^2s}+1}\right).
\end{eqnarray*}
Hence finally
\begin{equation}
\hat{F}^{+}(s,v)
=\frac{\sigma\sqrt{2}}{\sqrt{v^2+2\sigma^2u}+ v}.
\label{jf}
\end{equation}
With the identification $\sigma^2=2D$,
the result~(\ref{Flaplace}) is recovered.

For the binomial random walk,
the survival probability $F_n^+(v)$
is a highly non-trivial function of the velocity $v$,
depending on whether $v$ is rational or irrational,
because of the underlying lattice structure~\cite{bauer}.
In particular,
the limit survival probability $F^+(v)=\lim_{n\to\infty}F_n^+(v)$,
which is non-zero for $v<0$,
is discontinuous at any rational value of $v$.
The corresponding discontinuities are algebraic numbers
which can be determined explicitly~\cite{bauer}.
In the continuum limit ($v\to0^-$), $F^+(v)$
is simply given by $|v|\sqrt{2}/\sigma$.

In order to better understand the nature
of the corrections to the continuum limit,
we consider the simple case of the asymptotic mean occupation time
$\mean{T^+}$.
For the symmetric binomial random walk, and for fixed $v>0$, we have
\begin{equation}
\mean{T^+}=\sum_{n=1}^\infty\mean{I_n}=\sum_{n=1}^\infty2^{-n}
\sum_{k=k_0(v)}^n\bin{n}{k},
\label{jbin}
\end{equation}
with $k_0(v)=\integ(n(1+v)/2)+1$,
where $\integ(x)$ is the integer part of $x$,
i.e., the largest integer less than or equal to $x$.
As shown in the appendix, the behaviour of this expression as $v\to 0$ is
given by
\begin{equation}
\mean{T^+}=\frac{1}{2v^2}+\frac{A}{\sqrt{v}}+\frac{5}{12}+\cdots,
\label{jzeta}
\end{equation}
with
\[
A=\sqrt\frac{2}{\pi}\;\zeta\!\left(-\frac{1}{2}\right)=-0.165869209.
\]
The first term in~(\ref{jzeta})
corresponds to the continuum-limit result (see equation~(\ref{moment}))
\begin{equation}
\mean{T^+}_{\rm Brown}=\frac{1}{2v^2},
\label{jtasrw}
\end{equation}
because $D=1/2$ for the binomial random walk.
The second term in~(\ref{jzeta}) is surprising in several respects:
it is of relative order $v^{3/2}$, instead of the naturally expected $v^2$,
and the corresponding amplitude $A$ is transcendental.

Figure~1 shows a plot of the difference between the exact value
of $\mean{T^+}$, obtained by evaluating numerically~(\ref{jbin}),
and its continuum-limit expression~(\ref{jtasrw}).
The sum of the last two terms in~(\ref{jzeta}),
shown as a thick line, correctly describes the mean asymptotic behaviour
of the plotted quantity.
Superimposed periodic oscillations, with period two, are also clearly visible.
Similar oscillations, due to the lattice underlying the discrete walk,
are also encountered when considering other quantities~\cite{bauer}
(see especially Figures~14 and~15 therein).

\begin{figure}[htb]
\begin{center}
\includegraphics[angle=90,width=.8\linewidth]{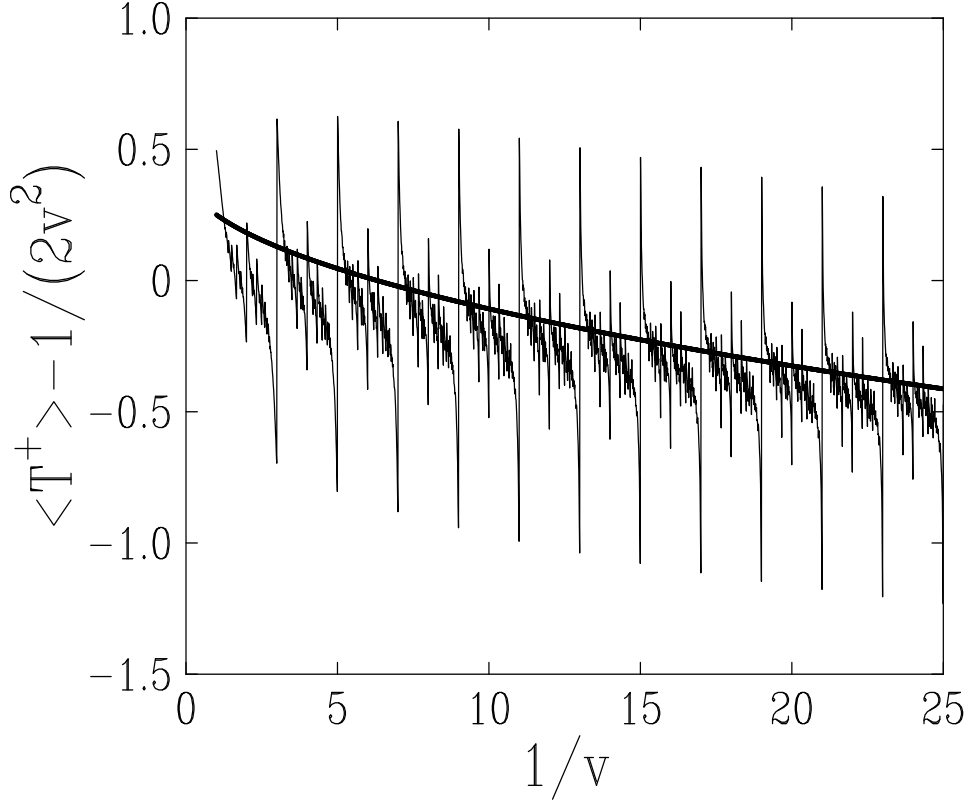}
\caption{\small
Plot of the difference between the mean occupation time $\mean{T^+}$
for the symmetric binomial random walk, and its continuum-limit
expression~(\ref{jtasrw}), against $1/v$.
Thick line: sum of last two terms of equation~(\ref{jzeta}).}
\end{center}
\end{figure}

Consider finally the case where the steps have a Cauchy distribution
\[
\rho(\eta)=\frac{1}{\pi(1+\eta^2)}.
\]
Because of the stability of the Cauchy law,
the probability $\mean{I_n}=\P(x_n>vn)$ is independent of $n$:
\begin{equation}
\mean{I_n}=\P(x_n>vn)=a(v)=\int_{v}^{\infty}\d u\,\rho(u)
=\frac{1}{2}-\frac{1}{\pi}\arctan v.
\label{cauch}
\end{equation}
Hereafter we denote this expression by $a$,
for short.
Using~(\ref{spar2}), we get
\[
\sum_{n=0}^{\infty}F_{n}^{+}(v)z^n=\left(1-z\right)^{-a},
\]
hence
\[
F_{n}^{+}(v)=\frac{\Gamma(n+a)}{n!\,\Gamma(a)}
\approx\frac{n^{-\theta(v)}}{\Gamma(a)}\qquad(n\gg1),
\]
with
\begin{equation}
\theta(v)=1-a=\frac{1}{2}+\frac{1}{\pi}\arctan v.
\label{tetav}
\end{equation}
The survival probability $F_{n}^{+}(v)$ falls off as a power law,
with a continuous family of persistence exponents
$\theta(v)$~\cite{dg,baldass,drg,dubna,dornic}.

From equation~(\ref{spar1}), we obtain
\begin{equation}
\P\left(T_{n}^{+}=k\right)
=\frac{\Gamma(a+k)}{k!\,\Gamma(a)}
\frac{\Gamma(n+1-k-a)}{(n-k)!\,\Gamma(1-a)}
=\bin{n}{k}\frac{B\left(a+k,1-a+n-k\right)}{B\left(a,1-a\right)},
\label{T+0}
\end{equation}
where the beta function is defined as
\[
B(a,b)=\int_{0}^{1}\d u\,u^{a-1}(1-u)^{b-1}=\frac{\Gamma\left(a\right)
\Gamma\left(b\right)}{\Gamma\left(a+b\right)}.
\]
Defining the $\beta$ density as
\[
\beta_{a,b}(x)=\frac{1}{B(a,b)}x^{a-1}(1-x)^{b-1}\qquad(0<x<1),
\]
one can rewrite~(\ref{T+0}) as
\begin{equation}
\P\left(T_{n}^{+}=k\right)=\int_{0}^{1}\d u\bin{n}{k}
u^{k}\left(1-u\right)^{n-k}\beta_{a,1-a}(u).
\label{T+2}
\end{equation}

In the continuum limit where $n$ and $k$ are simultaneously large,
with a fixed ratio $x=k/n$,
the binomial distribution inside~(\ref{T+2}) converges to $\delta(u-x)$,
so that the limiting probability density function of $T_n^+/n$ reads
\begin{equation}
\lim_{n\to\infty}f_{T_n^+/n}(x)=
\beta_{a,1-a}(x)=\frac{\sin\pi a}{\pi}x^{a-1}(1-x)^{-a}.
\label{arcsineg}
\end{equation}
In particular, if $v=0$, then $a=1/2$, and one recovers the arcsine
law~(\ref{arcsine}).

To conclude, let us briefly consider the case of a moving boundary
whose position obeys an arbitrary power law:
$X(t)=w\,t^\nu$~\cite{jb,jkr,bauer}.

If the steps have a narrow distribution,
the continuum description can again be used
in the regime of long times ($t\gg1$) and weak bias ($\abs{w}\ll1$).
A generalisation of~(\ref{jclt}) shows that the mean occupation time reads
\[
\mean{T_t^+}=\int_0^t\d t'\mean{I_{t'}}
=\frac{1}{2}\int_0^t\d t'\erfc\left(\frac{w}{\sigma\sqrt{2}}
(t')^{\nu-1/2}\right).
\]
The case $\nu=1/2$ therefore demarcates between two regimes.
For $\nu<1/2$, the bias $w$ is irrelevant,
so that $\mean{T_t^+}\approx t/2$,
and the arcsine law~(\ref{arcsine}) still holds asymptotically.
To the contrary, for $\nu>1/2$, any weak bias is relevant,
so that a non-trivial limiting law for the occupation time $T^+$
is expected~(for $w>0$), generalising~(\ref{jlim}),
with $\mean{T^+}\sim(\sigma/w)^{2/(2\nu-1)}$.
In the marginal situation of a parabolic boundary $(\nu=1/2)$,
there is a continuously varying persistence exponent $\theta(w)$~\cite{jb,jkr},
and the fraction $T^+/t$ admits a non-trivial limit distribution,
which also continuously depends on $w$.

If the steps have a symmetric broad (L\'evy) distribution,
with tails falling off as $\rho(\eta)\sim\abs{\eta}^{-\mu-1}$, with $0<\mu<2$,
the above discussion on the relevance of the bias $w$ still applies,
with the marginal situation being $\nu=1/\mu$.
The case of Cauchy-distributed steps in the presence
of a ballistic boundary $(\mu=\nu=1)$ is an interesting example
of this marginal situation,
where the dependence on the bias $w=v$ of the persistence
exponent~(\ref{tetav})
and of the limit distribution~(\ref{arcsineg}) are known explicitly.

\newpage

\section*{Appendix:~Expansion for $v\to0$ of expression~(\ref{jbin})}

\setcounter{equation}{0}
\def\theequation{A.\arabic{equation}}

In this appendix we investigate the behaviour as $v\to0$
of expression~(\ref{jbin}) of the mean occupation time $\mean{T^+}$, i.e.,
\[
\mean{T^+}=\sum_{n=1}^\infty\mean{I_n}=\sum_{n=1}^\infty2^{-n}
\sum_{k=k_0(v)}^n\bin{n}{k}.
\]
We set $u=1/v$, and introduce the Laplace transform
$L(s)=\mathrel{\mathop{\L}\limits_{u}}\mean{T^+}$.
The above expression yields
\[
L(s)=\frac{1}{s}\sum_{n=1}^\infty2^{-n}\sum_{k=k_0}^n\bin{n}{k}
\,\exp\left(-\frac{sn}{2k-n}\right),
\]
with $k_0=k_0(0)=\integ(n/2)+1$.

Introducing the contour-integral representation
\[
\bin{n}{k}=\oint\frac{\d z}{2\pi\ii}\,\frac{1}{z^{k+1}(1-z)^{n-k+1}},
\]
where the contour encircles the origin,
and summing over $k$ at fixed $\ell=2k-n\ge1$, we obtain
\[
L(s)=\frac{\e^{-s}}{s}\oint\frac{\d z}{2\pi\ii}
\sum_{\ell=1}^\infty\frac{4}{(2z)^\ell\Big(4z(1-z)-\e^{-2s/\ell}\Big)},
\]
hence, after some algebra,
\[
L(s)
=\frac{\e^{-s}}{s}\sum_{\ell=1}^\infty\frac{(1+W_\ell)^{-\ell}}{W_\ell}
=\frac{\e^s}{s}\sum_{\ell=1}^\infty\frac{(1-W_\ell)^\ell}{W_\ell},
\]
with $W_\ell=\sqrt{1-\e^{-2s/\ell}}$.

In the regime of interest ($s\to0$), we have
\[
L(s)=L_0(s)+L_1(s)+\cdots,
\]
with
\[
L_0(s)=\frac{1}{\sqrt{2s^3}}
\sum_{\ell=1}^\infty\sqrt{\ell}\,\e^{-\sqrt{2s\ell}},\qquad
L_1(s)=\sum_{\ell=1}^\infty
\left(\frac{1}{2\sqrt{2s\ell}}-\frac16\right)\e^{-\sqrt{2s\ell}},
\]
and so on.

The leading series $L_0(s)$ has to be investigated in some detail,
by means of its Mellin transform $M_0(x)$,
for which we obtain a closed-form expression:
\[
M_0(x)=\int_0^\infty\d s\,s^{x-1}\,L_0(s)=2^{2-x}\,\Gamma(2x-3)\,\zeta(x-2)
\qquad(\re x>3),
\]
where $\zeta$ is Riemann's zeta function.
The behaviour of the subleading series $L_1(s)$
can be estimated to leading order,
replacing the sum over $\ell$ by an integral over $\sqrt{2s\ell}$.
We thus obtain $L_1(s)\approx1/(3s)$.

Inverting successively the Mellin and Laplace transforms,
we obtain expression~(\ref{jzeta}), i.e.,
\[
\mean{T^+}=\frac{1}{2v^2}+\frac{A}{\sqrt{v}}+\frac{5}{12}+\cdots,
\]
with
\[
A=\sqrt\frac{2}{\pi}\;\zeta\!\left(-\frac{1}{2}\right)=-0.165869209,
\]
and where the dots stand for a contribution going to zero as $v\to 0$.

\end{document}